# Turbulent universal galactic Kolmogorov velocity cascade over 6 decades

Ka Ho Yuen*,1, Ka Wai Ho1, Chi Yan Law2,3, Avi Chen4, Alex Lazarian1,5,6

1Department of Astronomy, University of Wisconsin-Madison, USA

2Department of Earth and Space Sciences, Chalmers University of Technology, Sweden

3European Southern Observatory, Karl-Schwarzschild-Strasse 2, D-85748 Garching, Germany

4Department of Physics, City College, CUNY, USA

5Centro de Investigación en Astronomía, Universidad Bernardo O'Higgins, Santiago, General Gana 1760, 8370993, Chile

6Korea Astronomy and Space Science Institute, Daejeon 34055, Republic of Korea

*kyuen2@wisc.edu

**Summary Paragraph**

We report evidence for the existence of a universal, continuous turbulent cascade of velocity fluctuations with Kolmogorov -5/3 slope spanning 6 orders of length scales, from $10^4$ pc down to $10^{-2}$ pc. This was achieved by applying our innovative technique of separating density and velocity fluctuations to a set of spectroscopic surveys featuring various galactic spectral lines. This unified velocity cascade involves different interstellar phases from diffuse galactic media to dense self-gravitating clouds and persists despite interstellar phase transitions. The turbulent density fluctuations, on the other hand, do not show this universality. The value of the density spectral slope changes in different interstellar phases. This agrees with the expectation of compressible turbulence theory and demonstrates that the density is only an indirect tracer of interstellar turbulence. We report that the density fluctuations for clouds and filaments that are preferentially parallel to the magnetic field exhibit the spectral slope of –2. The universal grand turbulent velocity cascade that is established in our paper has significant implications for key galactic physical processes, including star formation, cosmic ray transport etc. We anticipate our result to be a starting point for in vitro models of multiphase interstellar turbulence studies with a significant impact for modeling of spiral galaxies.

**Introduction**

Interstellar turbulence is ubiquitous and plays a significant role in the formation and evolution of stars and molecular clouds[2-6], propagation and acceleration of cosmic rays[7], and heat and matter transport[6]. A fundamental issue critical to our understanding of interstellar turbulence is whether there exists a universal cascade that starts at the galactic injection scale and percolates all the interstellar media (ISM) phases, or whether different interstellar phases host their own turbulent cascades.

There are three central open questions regarding turbulent velocity fluctuations in the ISM. As stated by B. Rickett, one of the pioneers in our understanding of galactic turbulence, "*The **first** major question is whether we can infer the presence of turbulence from density observations alone.*"[59] Indeed, *plasma* density fluctuations can arise for various reasons not necessarily related to turbulence. For instance, power laws are observed as a result of properties of static objects, e.g., sizes of astrophysical dust grains[55]. The ***second*** major question is whether the turbulent cascade percolates all the galactic media or is only present in the warm media where the electron density fluctuations are measured. The ***third*** major question is whether the Kolmogorov velocity spectrum extends to scales larger than the injection scale, similar to what is observed in the Earth's atmosphere on the scales larger than the atmospheric scale height[54]. To answer these key questions, one must study the dynamics of motions that is imprinted in the velocity spectra in different galactic media, which is a direct tracer of Magneto-Hydrodynamic (MHD) turbulence[2-5].

Existing studies fail to determine whether the actual turbulent velocity fluctuations continuously follow Kolmogorov law across different phases of ISM. Evidence of the universal galactic cascade with the Kolmogorov[1] spectrum, which is frequently called the "Big Power Law in the Sky" [1-8], was presented in the 1990s and was based on the analysis of the electron density fluctuations[9] from $\sim 100 AU$ to $\sim 10^{-4} AU$. Subsequent works including H$\alpha$ measurements[10] and those recently by the Voyager mission[11] extended the Big Power Law from $\sim 30 pc$ to $10^{-4} AU$, although the new Voyager data shows a significant change of power law slope at the smallest scales. Nevertheless, the combined datasets[9-11] are consistent with the -5/3 Kolmogorov law and present, indeed, the largest spatial power law known in nature. Since these studies deal with density fluctuations, however, they are not fully compelling to many experts in the community as evidence of a continuous, multiscale turbulent cascade. Another high impact study of turbulent velocity dealt with the change of velocity dispersions of clouds as a function of their size resulted in the formulation of the famous Larson's Law[8,52,53]. Larson's Law is frequently quoted as evidence of the Kolmogorov turbulence in molecular clouds covering the scale of 0.1pc to 10pc. However, the spectral index relating the cloud size with the velocity varies from $v \sim l^{0.38}$ to $v \sim l^{0.5}$, which is different from the Kolmogorov's scaling $v \sim l^{1/3}$. Other measurements of the velocity spectra in molecular clouds[10,13] and HI 21cm lines[12,14] suffered from the effects of density contamination caused by thermal broadening which prevented the analysis of the continuity of the velocity spectra through different phases of the ISM.

**Velocity spectrum: Kolmogorov cascade for turbulent velocity fluctuations from galactic scales to star forming hubs**

This work utilizes the newly developed Velocity Decomposition Algorithm (VDA) approach based on the theory of PPV statistics[19,21] to an extended set of spectroscopic data that are publicly available (see **Table 1**). The VDA extracts and analyzes the pure turbulent velocity fluctuations in the form of "velocity caustics" ($p_v$). Velocity caustics result from crowding of moving emitters in the velocity space, which induce fluctuations of intensities in spectroscopic position-position-velocity (PPV) data cubes even when an incompressible turbulent fluid is constant in emitter density[19]. The relation between the statistics of observed caustics and the true three-dimensional velocity fluctuations is well established and well tested[19,21]. However, in the actual ISM the density fluctuations also contribute to the intensity fluctuations in the PPV space, and thermal broadening dilutes the effect of velocity caustics. Using the VDA we can separate the contributions of velocity and density fluctuation in PPV space and measure the thermal broadening effect on velocity caustics. The VDA provides a self-consistent reliable approach for processing the observational data for spectral lines corresponding to different ISM phases (**See Methodology**), including $H\alpha, HI$ and molecular tracers such as $^{12}CO, ^{13}CO, C^{18}O$.

Our work provides the first evidence of a continuous Kolmogorov cascade of velocity fluctuations for 6 orders of magnitude from the galactic scale to the star forming hub scale. For each spectroscopic velocity channel map, we select the peak channel that contains the maximal velocity caustics fluctuations and plot its normalized one-dimensional velocity spectrum $E_{caustics}(R)/Var(p_v)$ normalized by mean volume HI number density $n_H$ converted from the respective tracers (**Table 1**) as a function of the length scale $R$ in parsecs (pc). The corresponding results are presented in **Figure 1**. To plot the spectra in terms of their physical length scale, we select the ISM regions where we have accurate distance information. **Table 1** summarizes all parameters we have used in plotting **Figure 1**. Our "Big Power Law" illustrated in **Figure 1** spans $10^4$pc to $10^{-2}$pc. It ranges from the turbulence in the unstable neutral media (UNM) of a galactic arm (the Cattail)[22,60] down to star forming hub scale in NGC1333[23].

The presence of strong thermal broadening is indeed the most important reason why one cannot recover the Kolmogorov cascade in 21cm lines[21]. In contrast to previous studies, we have found that thermally broadened 21cm lines contain a significant contribution from density fluctuations, which could be non-Kolmogorov due to, for instance, turbulence driving and compressions[14,29,30]. The velocity fluctuations, which follow Kolmogorov cascade for almost all conditions[18], are being masked by the density fluctuations[19-21]. This masking is preventing a robust quantification of how turbulence is carried over in the multiphase ISM. By separating out the density and velocity contributions with the VDA technique, we can subtract out both the density contribution and the thermally broadening effect to reveal the true velocity fluctuations.

**Density spectrum: the –2 and –5/3 dichotomy for parallel and perpendicular clouds**

As mentioned earlier, the density spectra slope is not a robust measure of turbulence dynamics[21,29,36]. However, it carries important information about matter compressions and is essential for understanding many astrophysical processes such as star formation. In the case of subsonic turbulence, the density spectrum is expected to be a passive tracer of Kolmogorov turbulence and thus follows the Kolmogorov cascade[2-5]. This, in fact, explains the results of the earlier studies that reported the -5/3 spectral slope of the Big Power Law in the Sky obtained with the electron density fluctuations in warm ISM[9-11]. However, turbulence gets supersonic in colder phases of the ISM with density clumps produced by shocks[14,21] modifying the density spectrum. It has been consistently shown that the magnetized CNM HI emission maps have a density spectrum of -2[12,14], which is steeper than Kolmogorov scaling. In addition, effects of self-gravity[36] can affect the density spectral slope.

We can further analyze the properties of density fluctuations by distinguishing density structures that have different orientations with respect to the mean magnetic field direction[24-25]. The cloud-B offset records the relative orientation of the cloud and the mean magnetic field direction. Observational studies found that molecular clouds are preferentially either parallel or perpendicular to the mean magnetic field direction (hereafter "parallel clouds" and "perpendicular clouds"). Furthermore, it has been recently reported that the two types of clouds show different fragmentation and star formation properties[24-26]. In **Figure 2** we show a plot of the power spectral slope (PSS) for density and velocity fluctuations for each cloud listed in **Table 1** as a function of the cloud dynamic length scales (See **Table 2, Supplementary Figure 3**). While the spectrum of velocity fluctuations for all the clouds is consistent with -5/3 Kolmogorov law (*red circles,* see the **Methodology** for the justification), the density PSS are different for parallel and perpendicular clouds (*blue squares*). The cloud-B offset as a function of the cloud length scales is shown in **Figure 2** (*black crosses*). We can see from **Figure 2** that for parallel clouds the density PSS is roughly -2, but for perpendicular clouds the density PSS is significantly shallower and is approximately -5/3 (see the **Methodology**).

Furthermore, we also found that the density PSS is a function of length scale and is correlated with the B-cloud offset. Clouds that are larger than $0.56^{+0.54}_{-0.27}$pc, regardless of the tracer, have a density PSS closer to –2, while clouds smaller than this number have a density PSS much shallower than -2. Indeed, numerical calculations suggest that the PSS for the smaller clouds is approximately -5/3. Both the deviation of density PSS from -2 and the B-cloud offset happen concurrently at a length scale of the order of $\sim 0.5 pc$.

The transition of density PSS from preferentially -2 to preferentially -5/3 spans a length scale of $12.88^{+10.55}_{-5.80}$ pc (See **Supplementary Figure 2**) and is completed roughly also at the recently reported minimal transition scale for the CNM repo [52], which is indicated in **Figure 2** as the red dash line (0.1pc). Understanding the transitions of the density PSS in multi-phase media can help uncover the enigmatic processes of turbulent interstellar media.

**Consequences of the grand velocity power law**

The velocity power spectrum that is reported in this paper resolves the vital question of whether there exists a grand turbulent cascade that protrudes through the diffuse and dense molecular gases in the ISM of the Milky Way. It supports the physical picture that turbulence driven at large injection scale dominates the whole range of dynamics of the ISM. Our results show that additional sources of energy injection at small scales, e.g., due to self-gravity, may not significantly change the overall galactic turbulent cascade[21,36].

The velocity power spectrum that is reported in this paper also increases our understanding of how turbulence percolates throughout the entire ISM and answers the three questions posed in the introduction. We show (1) that density fluctuations are not a good tracer of velocity turbulence, (2) that there is exists a continuous Kolmogorov velocity from diffuse regions to dense clouds, and (3) that the largest turbulent length scales exceed the injection scale.

To answer the *first* major question of this paper, **Figure 2** in our work reestablishes that the plasma density fluctuations in large scales *should not* be a probe of the interstellar turbulent cascade. This echoes B. Rickett that "*Density perturbations are not directly involved in the (turbulence) dynamics.*"[59]. According to the theory of subsonic interstellar turbulence[4,5,29] both velocity and densities are expected to carry the same imprint from the turbulent cascade, i.e. they should both have Kolmogorov scaling. However, from **Figure 2** we see that the density fluctuations do not follow the Kolmogorov cascade at large scales (blue regions in **Figure 2**) where we expect the turbulence to be subsonic. A possible explanation for the inconsistency of theory and observation is that the combined effects of thermal broadening, radiative transfer of HI and the HI-$H_2$ conversion smear out the Kolmogorov scaling of subsonic phases in interstellar media.

To answer the *second* question of this paper, **Figure 1** demonstrates the velocity turbulent cascade is continuous and follows Kolmogorov law from diffuse regions to the dense molecular cloud gas. This fact is very interesting, since the galactic environment is different from isothermal incompressible fluid for which Kolmogorov scaling is theoretically established[1]. Moreover, the ISM undergoes several phase transitions[15] that include warm to cold gas and atomic to molecular hydrogen. In addition, a significant percentage of thermally Unstable Neutral Media (UNM)[12,15,16,49,51,60] exists with an equation of state and properties that are radically different from isothermal gas. It is not self-evident that the Kolmogorov law should cascade between these transitions and atypical phases.

To answer the *third* question of this paper, the largest length scale ($10^4$pc) that our spectrum covers in **Figure 1** significantly exceeds the expected injection scale of the turbulence ($10^2$pc). We expect that this can be a consequence of the inverse cascade that is observed, for example, in the turbulence of the Earth's atmosphere[54] where the Kolmogorov spectrum continues as 2D turbulence to the scales significantly larger than the atmospheric scale height. Further research should test to what extent this analogy is correct.

Furthermore, analyzing the amplitudes of the tracers' spectra in the Taurus region, we can also quantify how the turbulent energy is transferred when HI recombines to form molecular $H_2$ gases, and whether turbulence driving is caused by self-gravity. Our results demonstrate that during the transition, the amount of HI participating in the turbulent cascade decreases with the turbulent energy being carried to $H_2$ (red and brown spectra in **Figure 1,** see also the **Supplementary Figure 3**). We observe that the HI velocity spectrum in Taurus is cut off (0.3 pc, red line of **Figure 1**) but the turbulent cascade continues in $H_2$. We also observe in **Figure 1** a similar turbulent amplitude in molecular and diffuse gas. This suggests that self-gravity is not driving turbulence in these length scales since the amplitude of the turbulence spectra is not changing. This also shows that gravitational collapse in molecular clouds in the lengths scale studied here is not a major source of turbulence, a burning question that worries astronomers[57,58].

The universal grand turbulent cascade that exists in the ISM despite the phase transitions and density changes that take place within it has important implications for our understanding of the ISM. It suggests that the **ISM is a unified turbulent magnetized system.** This can guide the modeling of interstellar processes in the Milky Way and, by extension, in other spiral galaxies. It also has high impact implications for understanding key astrophysical processes in the ISM, including star formation, cosmic ray propagation and acceleration, transfer of energy and matter, and for predicting the galactic foreground for CMB studies. In particular, our work shows that the Kolmogorov cascade in the ISM also persists in the UNM phase, which is very helpful in modelling the formation of ubiquitous, anisotropic CNM as observed on the sky[49,51].

**Figures**

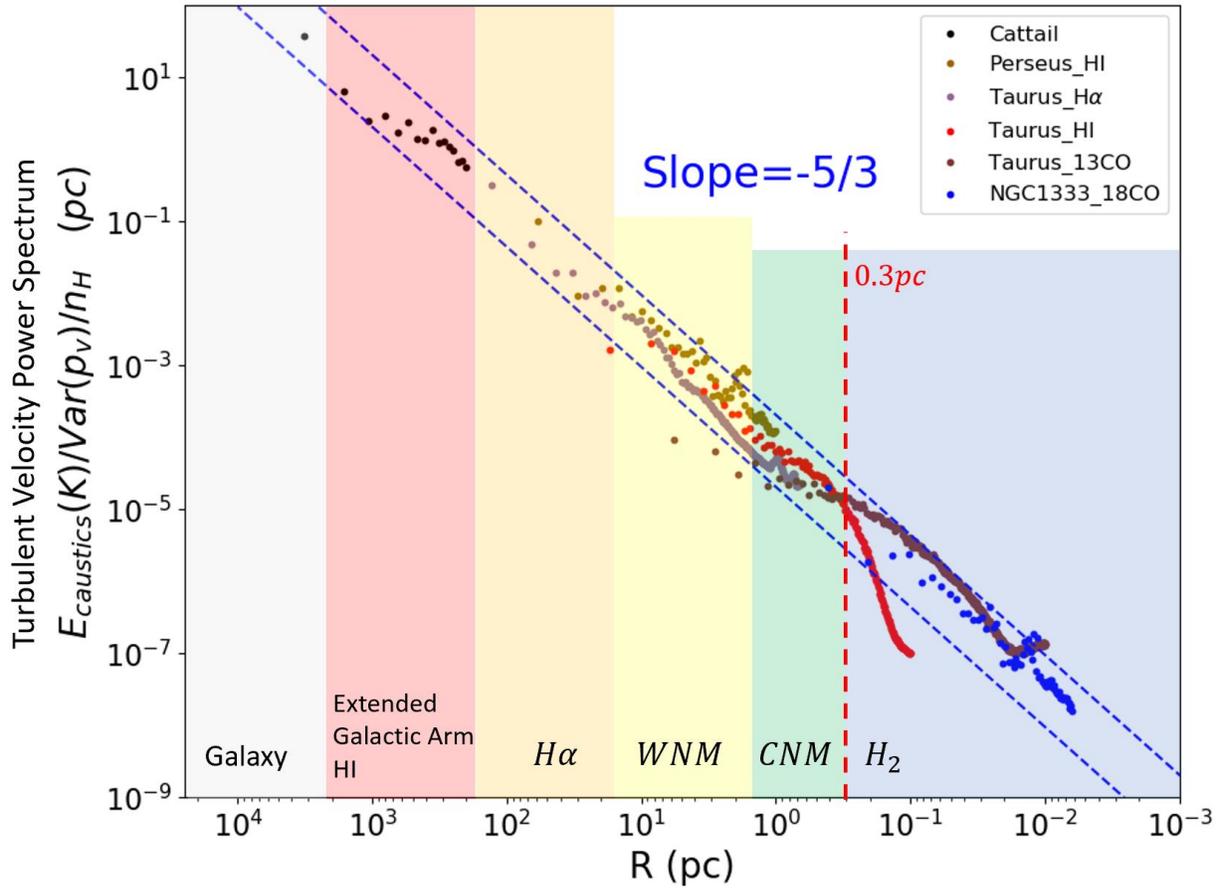

**Figure 1**: The Big Power Kolmogorov Law for ISM velocity fluctuations from $10^4$pc to $10^{-2}$pc using VDA[21] based on the data presented in **Table 1**. The continuous cascade is present for turbulent velocity from the galactic arm scale down to star forming hub. For reader's reference, we label the dominant scale of each phase. We also plot the auxiliary lines for Kolmogorov spectrum (the blue dash lines) to help the reader recognize the cascade.

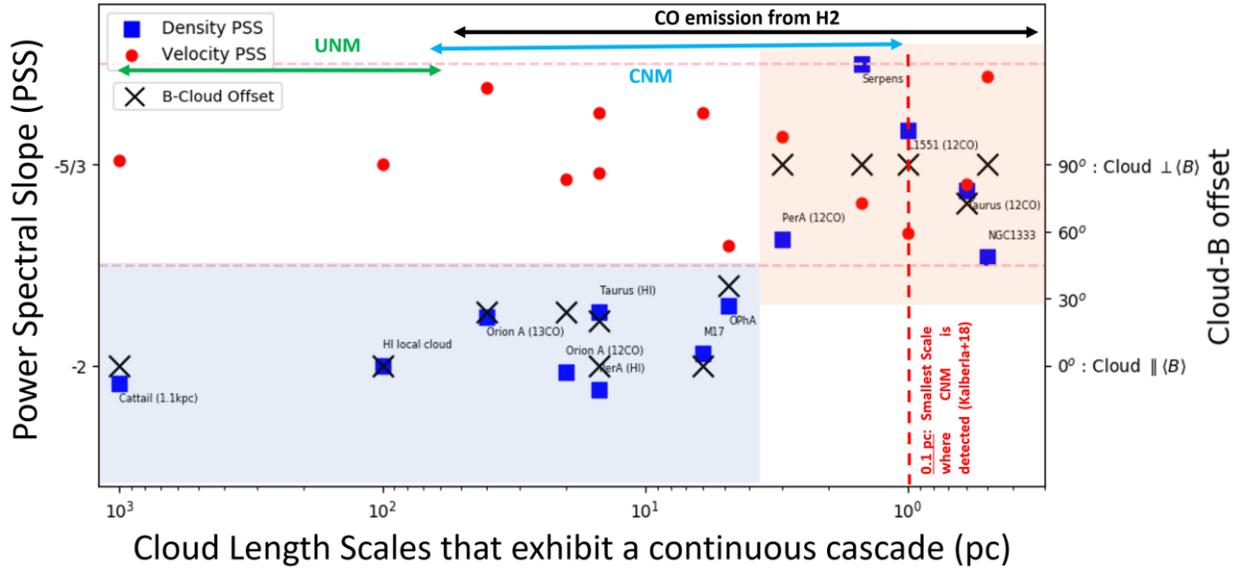

**Figure 2**: A plot showing the velocity (blue boxes) and density (red circles) power spectral slopes (PSS) as a function of cloud length scales that exhibit a continuous cascade (in pc) overplot with the Cloud-B offset (black crosses, see **Table 1**). We indicate the regions where the density PSS is preferentially -2 as shaded blue and -5/3 as shaded orange.

**Data Availability**

All the raw data that support the findings of this study are listed in Table 1, which the references and the direct links on where and how to download the data. To access the specific data, they are available from the corresponding author upon reasonable request.

**Code Availability**

We described the major steps in computing Fig 1 and Fig 2 in the supplementary materials. The main code for the computation of the spectrum is the VDA[21], where readers can find the corresponding Julia package in https://github.com/kyuen2/LazDDA

**Supplementary Information**

**Table 1: The essential information for the PPV data that we used in this study**

|  | Tracer | Line-of-Sight Distance ($pc$) | Cloud Major Axis Length ($pc$) | Tracible $n_H(cm^{-3})$ | Cloud-B offset (degrees) | Data Source |
|---|---|---|---|---|---|---|
| Cattail (1.1 kpc) | HI | 22000 | 1100 | 0.03 [22] | 0 | HI4PI[38] |
| OPhA | $^{13}CO$ | 140 | 4.8 | N/A | 36 [24,37] | Li et. al (2003)[48] |
| Taurus | $H\alpha$ | 140 | 15 | <5 [39] | 0 | WHAM DR1[39] |
|  | HI |  |  | 100 [40] | 0 | GALFA-DR2[40] |
|  | $^{12,13}CO$ |  |  | 28000 [47] | 73 [24,37] | Goldsmith et.al (2008)[42] |
| Perseus A | HI | 300 | 30 | 10 [40] | 20 [24,37] | GALFA-DR2[40] |
|  | $^{13}CO$ |  |  | N/A | 90 | Hatchell et. al (2005)[47] |
| Orion A | $^{13}CO$ | 350 | 40 | N/A | 24 [24,37] | Nakamura et. al (2019)[41] |
| M17 | $^{13}CO$ | 1600 | 6 | N/A | 0 | Nakamura et. al (2019)[41] |
| Serpens | $^{13}CO$ | 430 | 1.5 | N/A | 90 | Burleigh et. al (2013)[44] |
| NGC1333 | $C^{18}O$ | 235 | 0.5 | 50000 [43] | 90 | Curtis et.al (2010)[43] |
| L1551 | $^{13}CO$ | 140 | 1 | N/A | 90 | Lin et al (2016)[45] |

p.s. N/A means we did not use that in **Figure 1**. The respective links of the data can be found below:

**WHAM**: https://lambda.gsfc.nasa.gov/product/foreground/

**Cattail (HI4PI)**: http://cdsarc.u-strasbg.fr/viz-bin/qcat?J/A+A/594/

**GALFA-DR2**: Peek, Joshua, 2017, "GALFA-HI DR2 Narrow Data Cubes", https://doi.org/10.7910/DVN/MFM8C7, Harvard Dataverse, V1

**Taurus CO maps**: The original data is from Paul Goldsmith's 2008 paper. Data are available from the corresponding author upon having permissions from Paul.

**Perseus, OphA**: https://lweb.cfa.harvard.edu/COMPLETE/data_html_pages/FCRAO.html

**Orion A, M17**: http://jvo.nao.ac.jp/portal/nobeyama/sfp.do;jsessionid=F38D11999A00AB68AEDF91C25EFF0BDE

**Serpens**: http://cdsarc.u-strasbg.fr/viz-bin/cat/J/ApJS/209/39

**NGC1333**: https://doi:10.1111/j.1365-2966.2009.15658.x

**L1551**: https://www.nro.nao.ac.jp/~nro45mrt/html/data/L1551-13CO.fits.gz

**Table 2: The PSS results in terms of likelihood**

|  | Tracer | *Range of length scales R (pc)* | *Velocity PSS* | *Density PSS* |
|---|---|---|---|---|
| **Cattail (1.1 kpc)** | HI | 80-800 | $-1.66 \pm 0.05$ | $-2.03 \pm 0.11$ |
| **OPhA** | $^{13}CO$ | 1-10 | $-1.80 \pm 0.05$ | $-1.90 \pm 0.05$ |
| **Taurus** | $H\alpha$ | 10-100 | $-1.80 \pm 0.07$ | $-1.80 \pm 0.07$ |
|  | HI | 0.3-7 | $-1.58 \pm 0.02$ | $-1.91 \pm 0.02$ |
|  | $^{12}CO$ | 0.03-0.3 | $-1.70 \pm 0.02$ | $-1.71 \pm 0.03$ |
|  | $^{13}CO$ | 0.03-0.3 | $-1.33 \pm 0.01$ | $-1.33 \pm 0.01$ |
| **Perseus A** | HI | 2-80 | $-1.68 \pm 0.09$ | $-2.04 \pm 0.05$ |
|  | $^{12}$CO | 0.3-3 | $-1.62 \pm 0.02$ | $-1.79 \pm 0.02$ |
|  | $^{13}$CO | 0.3-3 | $-1.68 \pm 0.03$ | $-1.81 \pm 0.02$ |
|  | $^{13}$CO | 0.2-1 | $-1.82 \pm 0.04$ | $-1.95 \pm 0.04$ |
| **Orion A** | $^{12}CO$ | 1.6-4 | $-1.69 \pm 0.09$ | $-2.01 \pm 0.07$ |
|  | $^{13}CO$ | 4-40 | $-1.54 \pm 0.14$ | $-1.92 \pm 0.21$ |
| **M17** | $^{13}CO$ | 0.12-0.6 | $-1.58 \pm 0.05$ | $-1.98 \pm 0.05$ |
| **Serpens** | $^{13}CO$ | 0.019-0.038 | $-1.73 \pm 0.03$ | $-1.50 \pm 0.04$ |
| **NGC1333** | $C^{18}O$ | 0.01-0.1 | $-1.52 \pm 0.09$ | $-1.82 \pm 0.17$ |
| **L1551** | $^{12}CO$ | 9.3-28 | $-1.78 \pm 0.18$ | $-1.61 \pm 0.19$ |

**p.s. "Range of length scales"** means the length scales that exhibit a linear, continuous cascade. The range of scales are naturally smaller than that of the actual size of the cloud. See Supplementary Figure 3 for an example of Taurus' CO tracers.

**Acknowledgments**

K.H.Y., K.W.H & A.L. acknowledge the support from NSF AST 1816234, NASA TCAN 144AAG1967 and NASA ATP AAH7546. The numerical part of the research used resources from both Center for High Throughput Computing (CHTC) at the University of Wisconsin-Madison and National Energy Research Scientific Computing Center (NERSC), a U.S. Department of Energy Office of Science User Facility operated under Contract No. DE-AC02-05CH11231, as allocated by TCAN 144AAG1967. K.H.Y acknowledges the Mamba package from Julia (https://github.com/brian-j-smith/Mamba.jl) for the MCMC analysis.

**Author contributions**

Yuen & Ho, Lazarian designed the project. Yuen & Ho executed the project. All authors wrote up the manuscript. Yuen, Ho and Law analyzed the main results of the manuscript. Yuen and Ho establish the methodology of the paper. Yuen, Ho & Chen computed the Bayesian Analysis in the Methodology section and provided the analysis on large scale HI clouds. Yuen and Lazarian oversees the manuscript. All authors read and approve the manuscript.

**Methodology**

***Velocity Caustics from Velocity Decomposition Algorithm:*** Yuen et.al (2021) discussed that there is a special formula in reconstructing the velocity caustics $p_v$ from each channel based on a linear algebra formalism. This formalism is known to work in multiphase media. To summarize, the density fluctuation $p_d$ and velocity caustics $p_v$ can be obtained from the following formulae with the full PPV cube $p(\mathbf{R}, v)$ and the column intensity map $I(\mathbf{R})$:

$$p_v = p - (\langle pI \rangle - \langle p \rangle \langle I \rangle) \frac{I - \langle I \rangle}{\sigma_I^2}$$

$$p_d = (\langle pI \rangle - \langle p \rangle \langle I \rangle) \frac{I - \langle I \rangle}{\sigma_I^2}$$

For each piece of data (**Table 1**), we select the channel that has the maximal $p_v$ fluctuation in the PPV space.

***Normalized Power Spectral Slope:*** The 1D radial normalized power spectrum for any variable X can be obtained via the following formula

$$\frac{1}{Var(X)} E_X \left( K = \frac{1}{R} \right) = \frac{1}{Var(X)} \oint K d\Omega_K \, |\mathcal{F}\{X\}|^2$$

where $\Omega_K$ is the angular integral. To adjust the amplitude of the spectrum, we compute the quantity $\frac{E_X}{n_H Var(X)}$, where $n_H$ is the mean volume density of hydrogen *atom*. For molecular clouds, $n(H_2) = 2n_H$. The quantity $\frac{E_X(R)}{n_H Var(X)}$ has a unit of parsec.

*Product validity and error analysis of our result*:

**PSS = -5/3 for velocity fluctuations**: To produce **Figure 1**, we plot the quantity $\frac{E_X(R)}{n_H Var(X)}$ as a function of length scale R in parsec. The estimation of $n_H$ is given by literature provided in **Table 1**. The auxiliary lines in **Figure 1** are the 95% confidence interval of the spectral data point distributions. For artistic reason we select only a few spectra to be plotted in **Figure 1**. However, utilize the data points in **Table 2**, we can perform the Markov Chain Monte Carlo (MCMC) method to study the likelihood of the PSS of $p_v$ being $-5/3$ through the Julia package *Mamba*. We set up a linear probabilistic model assuming:

$$PSS_{p_v} = \beta_1 + \beta_2 R + dN$$

where we model both $\beta_{1,2}$ to follow normal distributions and $\sigma^2 = \langle dN^2 \rangle$. The interpretation for $\beta_1$ is the expected slope for PSS and $\beta_2$ is the dependence to length scales R. We run 3 rounds of simulations with 10000 iterations, and the first 250 of them are burn-in. **Supplementary Figure 1** shows the results of our MCMC simulations. We can see that the expected $\beta_1$ is indeed $\approx -5/3$ with a standard deviation of 0.03. Moreover, $\beta_2 \approx 0 \pm 10^{-3}$, indicating that the postulate of caustics PPS being $-5/3$ across the length scales that we studied is statistically very probable.

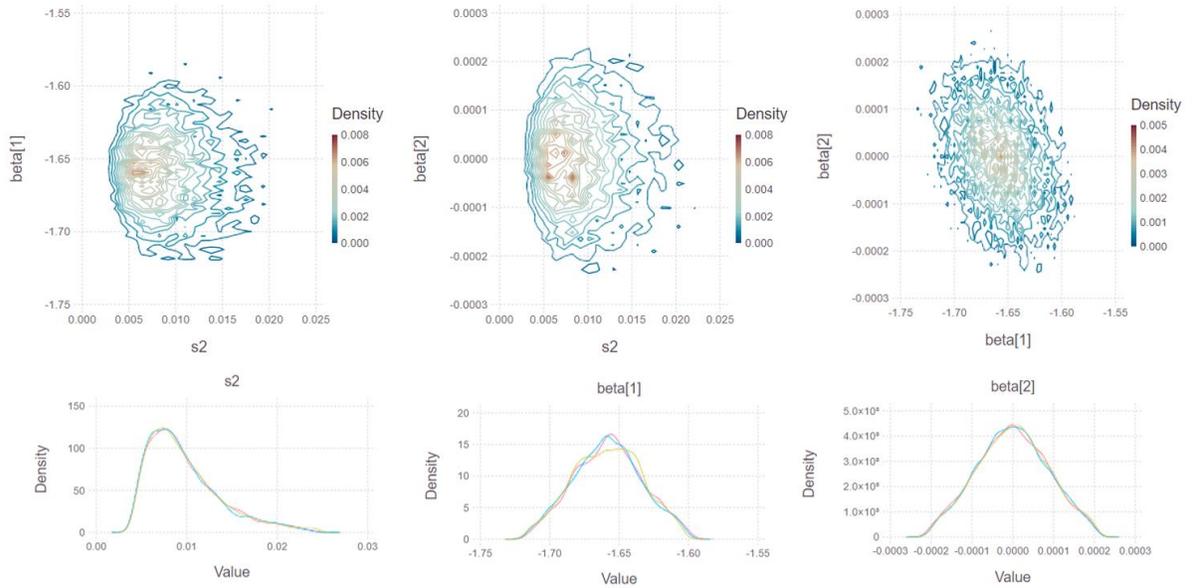

**Supplementary Figure 1**: The correlation plots (left column) and distributions (right column) of the results of our MCMC simulations for the three parameters $\sigma^2$ (s2), $\beta_1$ (beta[1]) and $\beta_2$ (beta[2]).

|  | MCMC result |
|---|---|
| $\beta_1$ (beta[1]) | $-1.66 \pm 0.03$ |
| $\beta_2$ (beta[2]) | $0.00000 \pm 0.0010$ |
| $\sigma^2$ (s2) | $0.010 \pm 0.005$ |

**Density PSS:** To produce the density PSS results in **Figure 2 and Table 2**, we perform a linear fit to each spectrum computed from publicly available data and classify whether they are preferentially $-2$, $-5/3$ or neither of them. To visually distinguish between whether the fitting data indeed falls into Kolmogorov cascade, we drew two pink lines representing the 10% deviation from the -5/3 line. From **Figure 2** we can see that all of the data points computed based on turbulent velocities fall into this 10% band. Notice that the linear fit from either spectrum or structure function is usually rather problematic due to unequal spectral or spatial binning and oscillations. For density fluctuations, it's only when $R \approx 0.56^{+0.54}_{-0.27}$pc will we see a transition of the slope from $\approx -2$ to $-5/3$.

To quantify the transition, we perform another MCMC simulation based on the following model

$$PSS_{density}(R) \sim \frac{5}{3} - \frac{1}{3} cdf(N(\log R + \mu, \sigma), R)$$

where $cdf$ is the cumulative distribution function, $N$ is the normal distribution function, and we expect $\mu[mu], \sigma[width]$ are stochastically normal. The MCMC result is summarized as **Supplementary Figure 2.** Notice that the expected transition from $-2$ to $-\frac{5}{3}$ is given by $10^\mu pc$, while the length scale width to transit from $-2$ to $-\frac{5}{3}$ is given by $10^\sigma pc$.

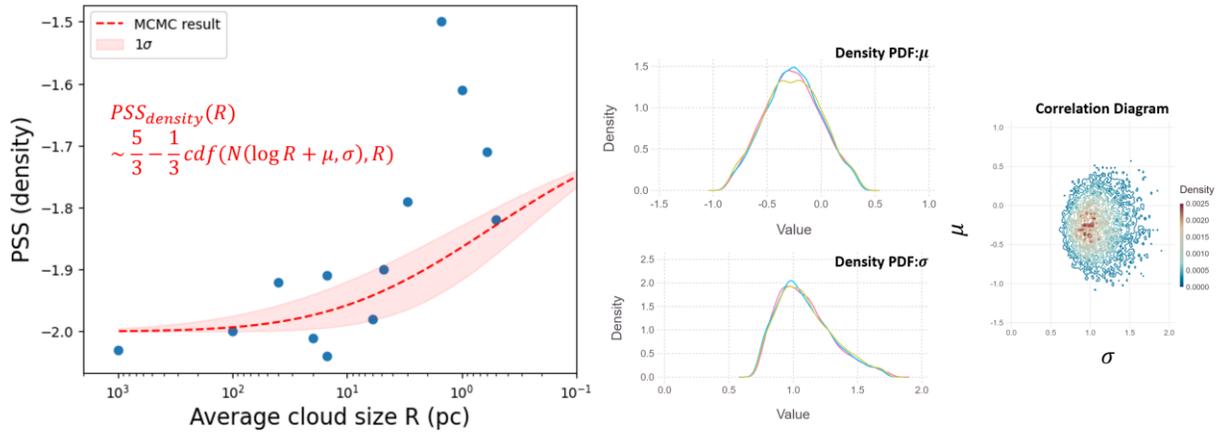

**Supplementary Figure 2.** (Left) The result of the MCMC simulations with probable bandwidth. (Top right) The correlation figure for $\mu$ (mu) and $\sigma$ (width). (Lower right) The Bayesian distributions for $\mu$ (mu) and $\sigma$ (width).

|  | MCMC result | Actual length scales (pc) |
| --- | --- | --- |
| $\mu$ (mu) | -0.26±0.30 | $0.56^{+0.54}_{-0.27}$ |
| $\sigma$ (width) | 1.11±0.26 | $12.88^{+10.55}_{-5.80}$ |

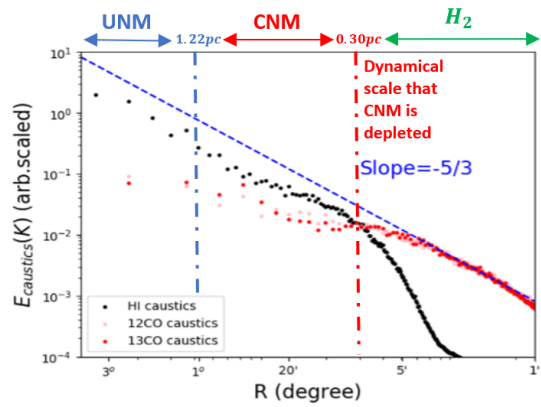

**Supplementary Figure 3.** The HI, 12CO and 13CO spectra for the Taurus region. The position R where the HI spectra deviates significantly from -5/3 is exactly when the CO spectra starts to follow -5/3 cascade. Notice that the "cloud size" traced by both the CO tracers are effectively up to ∼ 0.3pc since scales larger than $R > 0.3pc$ the spectrum exhibit non-linear oscillatory behavior.